\documentstyle[epsfig]{mn}

\title[Map making with data between 10 and 90~GHz]{Cosmic microwave background
map making with data between 10 and 90~GHz}
\author[A.W. Jones {\em et al.}]{A.W. Jones$^1$,  A.N. Lasenby$^1$,
P. Mukherjee$^1$,  C.M. Gutierrez$^3$, R.D. Davies$^2$, \cr R.A. Watson$^{2,3}$,
R. Hoyland$^3$ and R. Rebolo$^3$\\ 
  $^{1}$Mullard Radio Astronomy Observatory, Cavendish Laboratory, 
        Madingley Road, Cambridge CB3 OHE, UK\\
  $^{2}$University of Manchester, Nuffield Radio Astronomy 
        Laboratories, Jodrell Bank, Macclesfield, Cheshire, SK11 9DL, UK\\
  $^{3}$Instituto de Astrof\'\i sica de Canarias, 38200 La Laguna,
        Tenerife, Spain 
  }
\date{Accepted ???. Received ???; in original form \today}
\pagerange{\pageref{firstpage}--\pageref{lastpage}}
\pubyear{1999} 

\newcommand{\dg}{\nobreak^\circ }

\begin{document} 
\maketitle 
\label{firstpage}

\begin{abstract}
We use data from the Tenerife 10, 15 and 33~GHz beamswitching 
experiments along with the COBE 53 and 90~GHz data to separate the 
cosmic microwave background (CMB)
signal from the Galactic signal and create two maps at high 
Galactic latitude. The new multi-MEM technique is used to obtain the
best reconstruction of the two channels. The two maps are presented 
and known features are identified within each. We find that the
Galactic contribution to both the 15 and 33~GHz Tenerife data is
small enough to be ignored when compared to the errors in the data and
the magnitude of the CMB signal.
\end{abstract}


\section{Introduction}
\label{intro}

When analysing data from cosmic microwave background (CMB) experiments it is 
important to be able to distinguish features originating from the CMB and 
those that originate from foregrounds. With the sensitivity of experiments 
improving it is becoming increasingly important to subtract these foregrounds
from the data before any cosmological interpretation is made. Using 
multi-frequency observations it should be possible to extract information on 
both the CMB and foregrounds sources. 

The multi-maximum entropy method (multi-MEM) can be used to analyse 
multi-frequency observations to obtain constraints on the various 
foregrounds that affect CMB experiments. Hobson {\em et al.} (1998) use this 
method to analyse simulated data from the Planck satellite experiment 
and Jones {\em et al.} (1999a) apply it to simulated data from the MAP 
satellite. In 
applying this technique the different frequency observations must be 
sensitive to the same structures (i.e. they have an overlapping window 
function) as well as observing the same region. 

Fortunately the Tenerife beam-switching experiments have an overlapping 
window function and common observing region with the COBE satellite. This 
constitutes a region with a frequency range of 10~GHz 
to 90~GHz and, therefore, offers enough data to allow Galactic subtraction 
from CMB data. We present the results of this joint analysis in the form 
of a CMB map and Galactic foreground map at 10~GHz. 

\section{Tenerife data}
\label{tendat}

The observing strategy and telescope design of the Tenerife
experiments have been described
elsewhere (see Gutierrez {\em et al.} 1995 and references
therein). The data to be used in the analysis here have been presented
in a companion paper (Gutierrez {\em et al.} 1999) and consist of
scans at constant declination from a set of beam-switching
experiments. To make use of a large range in frequency coverage it
is only possible to use the data between Declinations $32.5\dg$ and
$42.5\dg$ where there is data at both 10~GHz and 15~GHz. The
Declination strips are separated by $2.5\dg$ and so there are 5 scans
in the final analysis (it is noted that the technique described below
does not require the same number of declinations at each frequency but
as we require a Galactic separation we are limited by our coverage at
the lowest frequency where the Galactic emission dominates). We also
concentrate our analysis at high Galactic latitudes away from strong
Galactic emission (RA $160\dg - 260\dg$) where the CMB signal will be
more dominant). The final
stacked data scans consist of continuous observations binned in $1\dg$
intervals in Right Ascension (RA) and the final average noise per bin
is $130\mu$K at 10~GHz and $54\mu$K at 15~GHz (corresponding to 
$50\mu$K and $20\mu$K per beam respectively). The beamwidths are
$4.9\dg$ and $5.2\dg$ FWHM at 10~GHz and 15~GHz, respectively. We also
include the 33~GHz ($5.0\dg$ FWHM and $55\mu$K error per bin), 
Dec. $40\dg$ Tenerife 
data that was presented in Hancock {\em et al.} (1994) as an extra
constraint. No other data at this frequency is currently available.

\section{COBE data}
\label{COBEdat}

The COBE satellite window function overlaps that of the Tenerife
window function (Watson {\em et al.} 1992) and so it should see the
same features. The sensitivity of the four-year COBE data is also
comparable to that of the Tenerife data and so it forms a useful
constraint on the level of CMB, relative to the Galactic emission,
contained in the two data sets. We extract the region overlapping that
observed by the Tenerife 10~GHz experiment at high Galactic latitude
(RA $160-260\dg$, Dec. $32.5-42.5\dg$) which corresponds to 200 COBE
pixels at each frequency. As the noise in the 30~GHz COBE channel is
much larger than that in the other two channels (53~GHz and 90~GHz)
and also much larger than in the Tenerife data we opt not to use this
data in the analysis. Our final data set therefore consists of data
covering RA $160-260\dg$, Dec. $32.5-42.5\dg$ at 10, 15, 33, 53 and
90~GHz covering almost an order of magnitude in frequency. 

\section{Multi-MEM analysis technique}
\label{mulitMEM}

The multi-MEM technique has been presented elsewhere (Hobson {\em et al.} 
1998). Here we will only outline the implementation of the technique to the
data set considered in this paper. As has been previously shown (Jones {\em 
et al.} 1998) the Tenerife data set contains long term baseline variations
which can be simultaneously subtracted from the data when applying the MEM 
technique. This is still done with the multi-MEM technique and the only 
difference between the single channel MEM presented in Jones {\em et al.} 
(1998) and the multi-MEM technique considered here is that the $\chi^2$ 
is now a sum over each of the frequency channels and the entropy is a sum
over each of the reconstructed maps (in this case the CMB and Galactic 
foreground maps). The information that the multi-MEM technique requires is 
only the frequency spectra of the channels to be reconstructed (although 
this can be relaxed if a search over spectral indices is performed as in 
Section~\ref{specfind}). The maximum entropy result was found using a
Newton-Raphson iteration method until convergence was obtained
(usually about 120 iterations were used although convergence was
reached within $\sim 60$ iterations).

\subsection{The choice of $\alpha$ and $m$}
\label{alphachoice}

\begin{figure}
\centerline{\hbox{\psfig{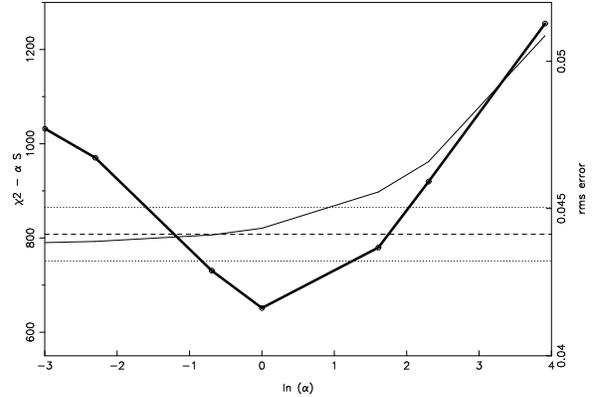}}}
\caption{The thin line shows 
$F=\chi^2 - \alpha S$ as a function of the Bayesian parameter
$\alpha$. The classical allowed range for $F$ is given between
the two dotted lines (see text). 
The thick line 
shows the error on the reconstruction as a function of $\alpha$. 
Each point was calculated over 50 Monte-Carlo simulations.}
\label{figa}
\end{figure}

\begin{figure}
\centerline{\hbox{\psfig{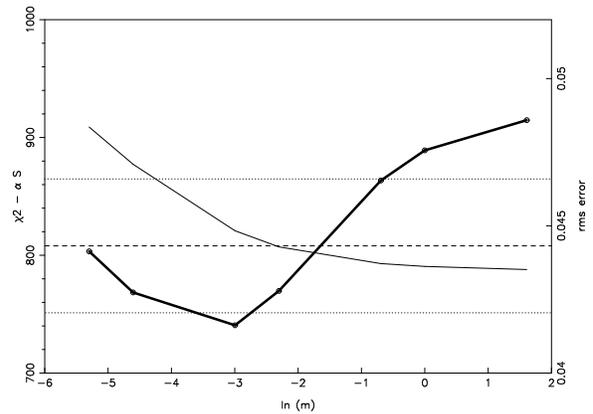}}}
\caption{The thin line shows 
$F=\chi^2 - \alpha S$ as a function of the model parameter
$m$. The classical allowed range for $F$ is given between
the two dotted lines (see text). 
The thick line
shows the error on the reconstruction as a function of $m$. 
Each point was calculated over 50 Monte-Carlo simulations.}
\label{figb}
\end{figure}

The choice of the Bayesian parameter $\alpha$ and the model parameter
$m$ in the Maximum Entropy method have often been treated as a
`guesswork'. In Fourier space it is possible to calculate the Bayesian
value for the $\alpha$ parameter but in real space this becomes much
more complicated as it involves the inversion of large matrices (see
Hobson {\em et al.} 1998). In the past $m$ was chosen to be the {\em
rms} of the signal expected in the reconstruction and $\alpha$ was
chosen so that the final value of $F=\chi^2 - \alpha S$ was
approximately equal to the number of data points, $N$. Actually $F$
was chosen to be just
below $N+2\sqrt{N}$ which is the confidence limit on $F$ calculated by
using the degrees of freedom on $\chi^2$ as a function of the
reconstruction. Here we investigate
the behaviour of the Monte-Carlo reconstructions with varying $\alpha$
and $m$ and show that this approximation is actually a very good
one. Figures~\ref{figa} and \ref{figb} show the variation of $F$ with
$\alpha$ and $m$ respectively. Also shown is the average error on the
reconstructions. As is seen the minimum in the errors on the
reconstruction for $\alpha$ occurs when $F$ is within the allowed
range for classic MEM ($F=N\pm 2\sqrt{N}$ where $N$ is the number of
data points). The minimum value
for $F$ when varying $m$ occurs where $m=50\mu$K which is the {\em
rms} of the expected reconstruction. Therefore, our initial `guesses' at
$\alpha$ and $m$ are properly justified and are the values used in the
following analysis. 

\section{Application to the data}
\label{appl}

\begin{figure}
\centerline{\hbox{\psfig{figure=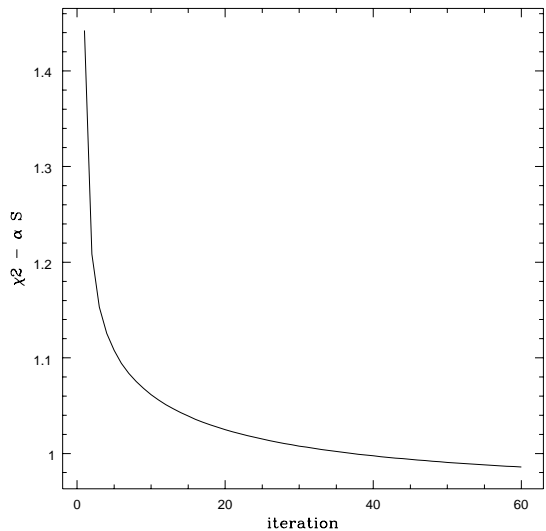,height=3.0in,angle=0}}}
\caption{$\chi^2 - \alpha S$ per pixel as a function of the number of
iterations of the Newton-Raphson minimisation process. It is seen that
the MEM algorithm converges very rapidly and approaches the classic
MEM result as required ($\chi^2 - \alpha S = N$ where $N$ is the number of
pixels).}
\label{figx}
\end{figure}

The multi-MEM technique was applied to the full Tenerife and COBE data set
assuming that there were two sources for the fluctuations seen in the data. 
The spectral dependencies of these two sources were that of CMB and that of
free-free emission (although this was later relaxed, see 
Section~\ref{specfind}). Simultaneous long term baseline removal was performed
on the Tenerife maps and any features with a period longer than $25\dg$ 
(corresponding to features that are outside the Tenerife window function) 
were subtracted. The convergence of the MEM algorithm is shown in
Figure~\ref{figx}. 

\begin{figure*}
\centerline{\hbox{\psfig{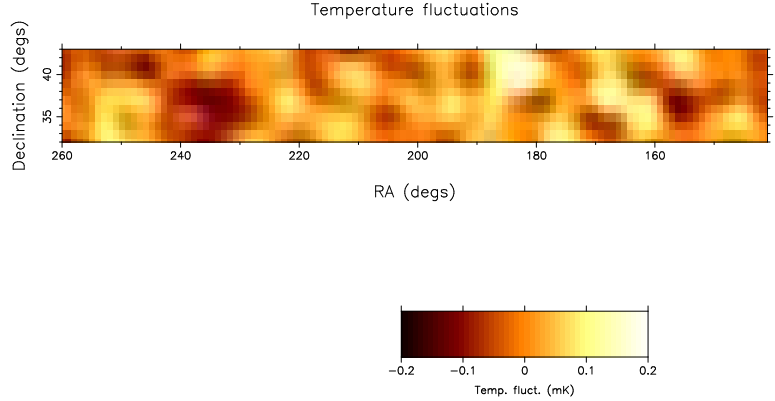}}}
\caption{The reconstructed CMB map at 10~GHz. The region shown here 
corresponds to the high Galactic latitude region used by the Tenerife
group as it is relatively low in Galactic contamination (at 10~GHz the 
signal from the CMB and Galaxy are roughly equal on $5\dg$
scales). The error on this reconstruction is $40\mu$K per $1\dg$
square pixel or $9\mu$K per $5\dg$ feature.}
\label{fig1}
\end{figure*}

\begin{figure*}
\centerline{\hbox{\psfig{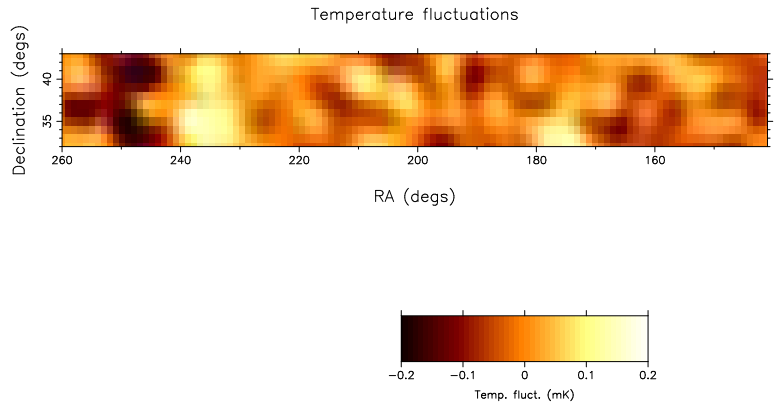}}}
\caption{The reconstructed Galactic map at 10~GHz. The region is the same
as Figure~\ref{fig1}. The error on this reconstruction is $30\mu$K per
$1\dg$ square pixel or $7\mu$K per $5\dg$ feature.}
\label{fig2}
\end{figure*}

Figure~\ref{fig1} shows the CMB reconstruction obtained and Figure~\ref{fig2}
shows the Galactic foreground reconstruction. As can be seen the two maps
clearly have different features. The sky coverage is limited by the smallest 
survey (10~GHz) although the single declination at 33~GHz (Dec. $40\dg$) is 
included in the analysis as an extra constraint. The {\em rms} signal
at 10~GHZ for the CMB and Galactic maps are $42\mu$K and $36\mu$K respectively.
Taking a free-free spectral index this corresponds to 
a Galactic signal of $15\mu$K at 15~GHz and $3\mu$K at 33~GHz which
implies that the Galactic foreground is negligable in both cases (when
added in quadrature). The error on the CMB
and Galactic reconstruction is $40\mu$K and $30\mu$K per $1\dg$
pixel respectively (compared with errors of $46\mu$K and $35\mu$K 
on the reconstructions of the separate 10~GHz and 15~GHz channels respectively 
presented in Gutierrez {\em et al.} 1999). This error was calculated from the variance 
over 300 Monte-Carlo simulations of the MEM analysis. The
reconstruction contains individual features at $5\dg$ resolution (the
maximum resolution of the Tenerife experiments) and the error over one
of these features is therefore $\sim 9\mu$K for the CMB map and $\sim
7\mu$K for the Galactic map.

\subsection{Spectral index determination}
\label{specfind}

\begin{figure}
\centerline{\hbox{\psfig{figure=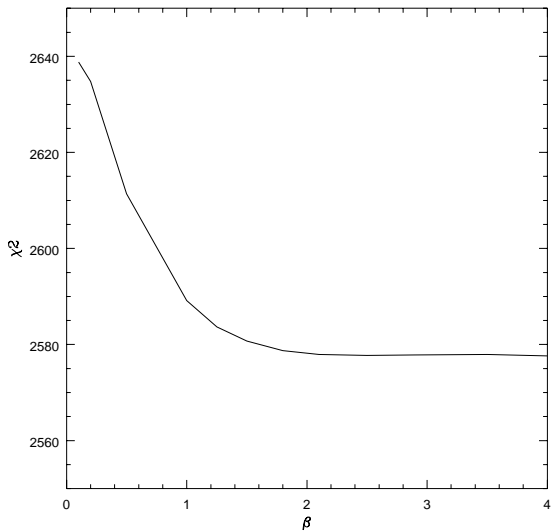,height=3.0in,angle=0}}}
\caption{$\chi^2$ as a function of the input spectral index ($T
\propto  \nu^{-\beta}$) of the Galactic 
channel. A spectral index of 2.1 corresponds to free-free emission.}
\label{fig3}
\end{figure}

The analysis performed assumed that the Galactic foreground was free-free
emission with a temperature spectral index of -2.1. Clearly, this is an 
assumption which may be incorrect as the dominant foreground in this region 
could be from another source. Therefore, it is necessary to vary this 
spectral index to see how the foreground and CMB reconstructions change. 
This has been done over a range for the spectral index 
of 0.0 to -4.0. Figure~\ref{fig3} shows the
variation in $\chi^2$ (comparing the predicted data to the real data) 
over this range of spectral index. As can be seen there is a broad minimum 
above a spectral index of $\sim 2$ which implies that in this range very little
is changed in the two reconstructions. This is because the main information
on the Galactic channel occurs in the 10~GHz data and the higher frequency
data acts as upper limits. Therefore, it appears as though the Galactic 
foreground fits the data well with a spectral index of $\sim 2$ which 
corresponds to free-free emission (this spectral index corresponds to the
point at which the Galactic contribution at the higher frequencies is just
below the upper limit). Changes in the CMB reconstruction were well
below the noise when the spectral index was varied for the Galactic channel.

\section{Identification of features}
\label{features}

The main purpose of using the multi-MEM technique is to allow clean separation
of CMB sources from foreground emission. This can be checked by comparing
the two maps with existing surveys and previous predictions. 

It is very difficult to compare the features in the maps with known 
Galactic features as very few surveys at similar frequencies and scale 
cover the Tenerife region of sky. However, there is a survey at 5~GHz which
uses an interferometer and has been used to create a map of Galactic 
fluctuations covering this region (Jones {\em et al.} 1999b, hereafter JB98). 
Also, the 408~MHz (Haslam, Salter, Stoffel \& Wilson 1982)
and 1420~MHz (Reich \& Reich 1988) 
surveys cover this region but the artefacts within the 1420~MHz
survey makes a meaningful comparison difficult (JB98). By 
comparison with the survey presented in JB98 some common features are observed.
For example, the Galactic feature at RA $175\dg$, Dec. $32.5\dg$ was seen
in the 5~GHz and the 408~MHz survey. The point source at Dec. $30\dg$, 
RA $200\dg$ appeared to be extended in the 5~GHz survey and there appears to 
be a Galactic feature in the same region which could account for this 
extension. However, there are features which appear in the Galactic map
here and not in the lower frequency surveys although this could be due to 
the different angular scale dependencies. For example, there is a large
feature at RA $230\dg$, Dec. $35\dg - 40\dg$. This
only shows up in the 5~GHz map at very small amplitude and this could be 
due to the interferometer resolving out the feature. This feature was
observed at Dec. $40\dg$ in an $8\dg$ FWHM beam at 10~GHz using the
predecessor to the Tenerife experiments (Davies {\em et al} 1987). 

The CMB map is even more difficult to compare with other maps as the only 
two surveys which have covered this region are the COBE and Tenerife 
experiments. Therefore, no comparison is possible at this time although it
is possible to check previous predictions made by the COBE and Tenerife
teams. The main predictions about the CMB were the feature at Dec. $40\dg$, 
RA $185\dg$ (Hancock {\em et al.} 1994) and the positive-negative-positive 
feature (smoothed to $10\dg$ scale in RA) at 
Dec. $35\dg$, RA. $200-250\dg$ (Bunn, Hoffman \& Silk, 1996 
and Gutierrez {\em et al.} 1997), both of which are clearly seen here
(the second being a combination of negative and positive fluctuations). The
other features in this map are all potentially CMB in origin but must 
wait until other experiments with overlapping window functions have 
surveyed this region before being assigned unambiguously. 

\section{Conclusions and further work}
\label{conc}

The reconstructed maps using data between 10~GHz and 90~GHz were
presented. Features of Galactic and cosmological origin were
identified. The CMB map is very similar to the reconstructed map using
the 15~GHz Tenerife data alone (Gutierrez {\em et al.} 1999) and so it
is possible to use that data alone as a constraint on the CMB to put
limits on the cosmological parameters as has
been done in the past. It was found that the Galactic contamination in
this frequency range is consistent with free-free emission (if the
upper limit set by the 15~GHz data is a true limit), although no
constraint on the relative level of free-free and synchrotron emission
was possible. Taking all of the Galactic foreground to be free-free
emission it was found that the 15~GHz data was contaminated by
$15\pm 3\mu$K which is very small when added in quadrature to the CMB
signal of $42\pm 9\mu$K. If the foreground was all synchrotron emission then this
value would be even lower.  At 33~GHz the free-free component would
contribute a signal of $3.0\pm 0.6\mu$K which is negligible.

There are many future applications of this technique. The multi-MEM
has been applied to simulated Planck and MAP satellite data
already. We are presently collating other CMB and Galactic surveys
together to put further constraints on the foregrounds and the CMB
itself at different angular scales and covering a wider frequency
range to reduce the errors obtained here. The only constraint that
this method has on producing real CMB maps is that there is enough
frequency coverage of experiments with over lapping window functions
that observe the same region of sky. We are also
combining data from experiments with different window functions to put
direct constraints on the spatial power spectrum of the CMB and
Galactic fluctuations. We are currently working on applying this
technique to the spherical sky and a full likelihood analysis, as well
as tests for non-Gaussianity within the data, will be presented soon.

\section*{Acknowledgements}
 
AWJ acknowledges King's College, Cambridge,
for support in the form of a Research Fellowship.

\label{lastpage} 

\end{document}